\documentclass[twocolumn]{aastex631}
\usepackage{graphicx}
\usepackage{epstopdf}
\usepackage{float}
\usepackage{ulem}

\usepackage{color}
\usepackage{soul}

\renewcommand\baselinestretch{1.3}

\shortauthors{Fang et al.}

\begin{document}
\title{Automatic Classification of Galaxy Morphology: A Rotationally-invariant supervised Machine-learning method based on the Unsupervised Machine-learning Data Set}

\author[0000-0001-9694-2171]{GuanWen Fang}
\affil{Institute of Astronomy and Astrophysics, Anqing Normal University, Anqing 246133, People's Republic of China; \url{wen@mail.ustc.edu.cn}}

\author{Shuo Ba}
\altaffiliation{Shuo Ba and GuanWen Fang contributed equally to this work}
\affil{School of Engineering, Dali University, Dali 671003, People's Republic of China; \url{zhouchichun@dali.edu.cn}}

\author[0000-0003-3196-7938]{Yizhou Gu}
\affil{School of Physics and Astronomy, Shanghai Jiao Tong University, 800 Dongchuan Road, Minhang, Shanghai 200240, People's Republic of China; \url{guyizhou@sjtu.edu.cn} }

\author[0000-0001-8078-3428]{Zesen Lin}
\affil{CAS Key Laboratory for Research in Galaxies and Cosmology, Department of Astronomy, University of Science and Technology of China, Hefei 230026, People's Republic of China; \url{xkong@ustc.edu.cn}}
\affil{School of Astronomy and Space Science, University of Science and Technology of China, Hefei 230026, People's Republic of China}
\affil{Department of Physics, The Chinese University of Hong Kong, Shatin, N.T., Hong Kong S.A.R., China}

\author{Yuejie Hou}
\affil{School of Engineering, Dali University, Dali 671003, People's Republic of China; \url{zhouchichun@dali.edu.cn}}

\author{Chenxin Qin}
\affil{School of Engineering, Dali University, Dali 671003, People's Republic of China; \url{zhouchichun@dali.edu.cn}}

\author[0000-0002-5133-2668]{Chichun Zhou}
\affil{School of Engineering, Dali University, Dali 671003, People's Republic of China; \url{zhouchichun@dali.edu.cn}}

\author{Jun Xu}
\affil{Institute of Astronomy and Astrophysics, Anqing Normal University, Anqing 246133, People's Republic of China; \url{wen@mail.ustc.edu.cn}}

\author{Yao Dai}
\affil{Institute of Astronomy and Astrophysics, Anqing Normal University, Anqing 246133, People's Republic of China; \url{wen@mail.ustc.edu.cn}}

\author{Jie Song}
\affil{CAS Key Laboratory for Research in Galaxies and Cosmology, Department of Astronomy, University of Science and Technology of China, Hefei 230026, People's Republic of China; \url{xkong@ustc.edu.cn}}
\affil{School of Astronomy and Space Science, University of Science and Technology of China, Hefei 230026, People's Republic of China}

\author[0000-0002-7660-2273]{Xu Kong}
\affil{CAS Key Laboratory for Research in Galaxies and Cosmology, Department of Astronomy, University of Science and Technology of China, Hefei 230026, People's Republic of China; \url{xkong@ustc.edu.cn}}
\affil{School of Astronomy and Space Science, University of Science and Technology of China, Hefei 230026, People's Republic of China}

\noaffiliation



\begin{abstract}
Classification of galaxy morphology is a challenging but meaningful task for the enormous amount of data produced by the next-generation telescope.
By introducing the adaptive polar coordinate transformation, we develop a rotationally invariant supervised machine learning (SML) method that ensures consistent classifications when rotating galaxy images, which is always required to be satisfied physically but difficult to achieve algorithmically.
The adaptive polar coordinate transformation, compared with the conventional method of data augmentation by including additional rotated images in the training set, is proved to be an effective and efficient method in improving the robustness of the SML methods.
In the previous work, we generated a catalog of galaxies with well-classified morphologies via our developed unsupervised machine learning (UML) method. By using this UML-dataset as the training set, we apply the new method to classify galaxies into five categories (unclassifiable, irregulars, late-type disks, early-type disks, and spheroids). In general, the result of our morphological classifications following the sequence from irregulars to spheroids agrees well with the expected trends of other galaxy properties, including S\'{e}rsic indices, effective radii, nonparametric statistics, and colors. Thus, we demonstrate that the rotationally invariant SML method, together with the previously developed UML method, completes the entire task of automatic classification of galaxy morphology.

\end{abstract}


\section{Introduction}
Galaxy morphology is closely related to the formation and assembly history of the galaxy.
For example, as galaxies evolve, structures, such as bulges, bars, and spiral arms or tidal tails, are formed.
Moreover, morphology is related to color, stellar mass, star formation rate, gas content, and environments (e.g., \citealt{1980ApJ...236..351D, 2003MNRAS.341...54K, 2004ApJ...600..681B, 2019A&A...631A..38L}).
Therefore, the classification or quantification of galaxy morphology is instrumental in understanding galaxy evolution.
Besides the model-dependent or model-independent parameters developed to quantify galaxy morphology or structural features \citep{1963BAAA....6...41S, 2004AJ....128..163L, 2014ARA&A..52..291C}, morphological classification of galaxies \citep{1926ApJ....64..321H} is a fundamental problem and attracts great interests.

A direct way to carry out a certain scheme of the morphological classification is by visual
inspections  (e.g., \citealt{1926ApJ....64..321H, 1959HDP....53..275D, 1960ApJ...131..558V}).
For example, volunteers were gathered to identify the morphological types of galaxies in the Galaxy Zoo project (e.g.,  \citealt{2011MNRAS.410..166L, 2017MNRAS.464.4420S, 2022MNRAS.509.3966W}).
Besides the visual inspections, methods based on the multi-dimensional morphological
parameter space are developed, where the boundary of distinctions
can be defined by the empirical cuts \citep{2004AJ....128..163L,2014ARA&A..52..291C}
or machine learning algorithms, such as the principal component analysis \citep{2007ApJS..172..406S} and the support-vector
machine \citep{2008A&A...478..971H}.

In recent years, deep learning methods are introduced to the morphological classification of galaxies.
Instead of searching the boundaries in the parameter spaces consisting of manually designed parameters, they directly extract key features from the two-dimensional images with single or multiple channels and give the morphological types. For example, the convolutional neural network (CNN) can directly extract enormous information from raw pixels hierarchically and is capable of mimicking human perceptions.
The supervised neural networks had been applied to galaxy classification in several imaging surveys, e.g., Sloan Digital Sky Survey, Dark Energy Survey, and Cosmic Assembly Near-infrared Deep Extragalactic Legacy Survey at higher redshift \citep{2015MNRAS.450.1441D, 2015ApJS..221....8H, 2018MNRAS.476.3661D, 2019MNRAS.484...93D, 2020A&C....3000334B, 2020MNRAS.493.4209C, 2021MNRAS.507.4425C, 2021MNRAS.506.1927V}. Beyond the morphological classification, they were also used to identify galaxies with specific features such as bar structures \citep{2018MNRAS.477..894A} or galaxies that might suffer gravitational lensing effect \citep{2020ApJ...899...30L}.

However, the existing solutions of galaxy morphological classification suffer from the following shortcomings. (1) The supervised deep learning methods require large pre-labeled data sets as the training sets. Usually, such data sets are obtained by visual inspections, which are low efficiency, high cost, and with subjective bias.
(2) The CNN-based neural networks have poor robustness to image rotations (see \citealt{cheng2016learning,cheng2018learning,cabrera2017deep,chen2018rotation,reyes2018enhanced,yao2019rotation}). That is, the algorithms might misclassify the morphological types of galaxies after rotating their images.
Usually, the conventional solution to this problem is to perform data augmentation by including additional rotated images in the training set (e.g., \citealt{2015MNRAS.450.1441D}), which inevitably consumes a huge amount of unnecessary computing resources.

The unsupervised techniques that combine a convolutional autoencoder (CAE; \citealt{masci2011stacked}) with clustering methods have been applied in the classification tasks, such as gravitational lensing detection \citep{2020MNRAS.494.3750C} and galaxy morphological classification \citep{2021MNRAS.503.4446C}. The CAE is used to extract key features of galaxies from the raw images, while clustering algorithms are responsible for gathering the galaxies with similar features into a group subsequently. Different clustering algorithms, using different similarity definitions and techniques, may result in inconsistent clustering output. To obtain a comprehensive perspective of classifications, we develop the bagging-based multi-clustering method that clusters galaxies by the voting of three clustering algorithms \citep{2022AJ....163...86Z}, rather than using a single algorithm. To establish a high-quality classification of galaxy morphology, only the galaxies with consistent voting by the three clustering algorithms are collected as the ``well-classified'' dataset (hereafter the UML-dataset). In \cite{2022AJ....163...86Z}, the application of the proposed UML method to $47,149$ galaxies with $H < 24.5$ in five CANDELS fields resulted in a UML-dataset containing $24,900$ galaxies ($\sim 53\%$), at the cost of leaving $22,249$ galaxies ($\sim 47\%$) with disputed labels.

In this work, to handle the remaining $22,249$ galaxies with disputed labels and complete the entire task of automatic classification of galaxy morphology, we further propose a rotationally invariant supervised machine learning (SML) method by using the UML-dataset as the training set.
Unlike the conventional method of data augmentation, we propose a different method (adaptive polar coordinate transformation; APCT) to consider the rotation invariance of the CNN model in the pre-processing step, which converts the rotation-invariant problem into a translation-invariant problem.

The paper is organized as follows. The sample selection and UML-dataset are described in Section~\ref{sec:data}. The APCT and other pre-process strategies are introduced in Section~\ref{sec:data_pre}. In Section~\ref{sec:method}, we introduce three typical existing SML models. The combination of the APCT and the SML algorithms is used directly to give the morphological types of galaxies.
In Section~\ref{sec:result}, the result of morphological classification is given.
The effectiveness of the proposed method is evaluated by the t-SNE visualization graphs and galaxy properties as a function of our classification results from the best model. Finally, a summary is given in Section~\ref{sec:sum_outlooks}.

\section{Sample selection and UML-dataset}
\label{sec:data}

In this section, we give a brief introduction to the galaxy sample, which is in line with \cite{2022AJ....163...86Z}. The sample of $47,149$ galaxies with F160W $<$ 24.5 mag are selected from the five CANDELS fields, with an additional criterion being the flag {\tt use\_{phot}=1}, which means that the object is not a star and not heavily contaminated. Here we refer to the 3D-HST project \citep{2014ApJS..214...24S, 2016ApJS..225...27M} for the full details, from which the H-band selected catalogs (v4.1.5) and H-band images are taken in this work.

\begin{figure*}[htb!]
\centering
\includegraphics[width=\textwidth]{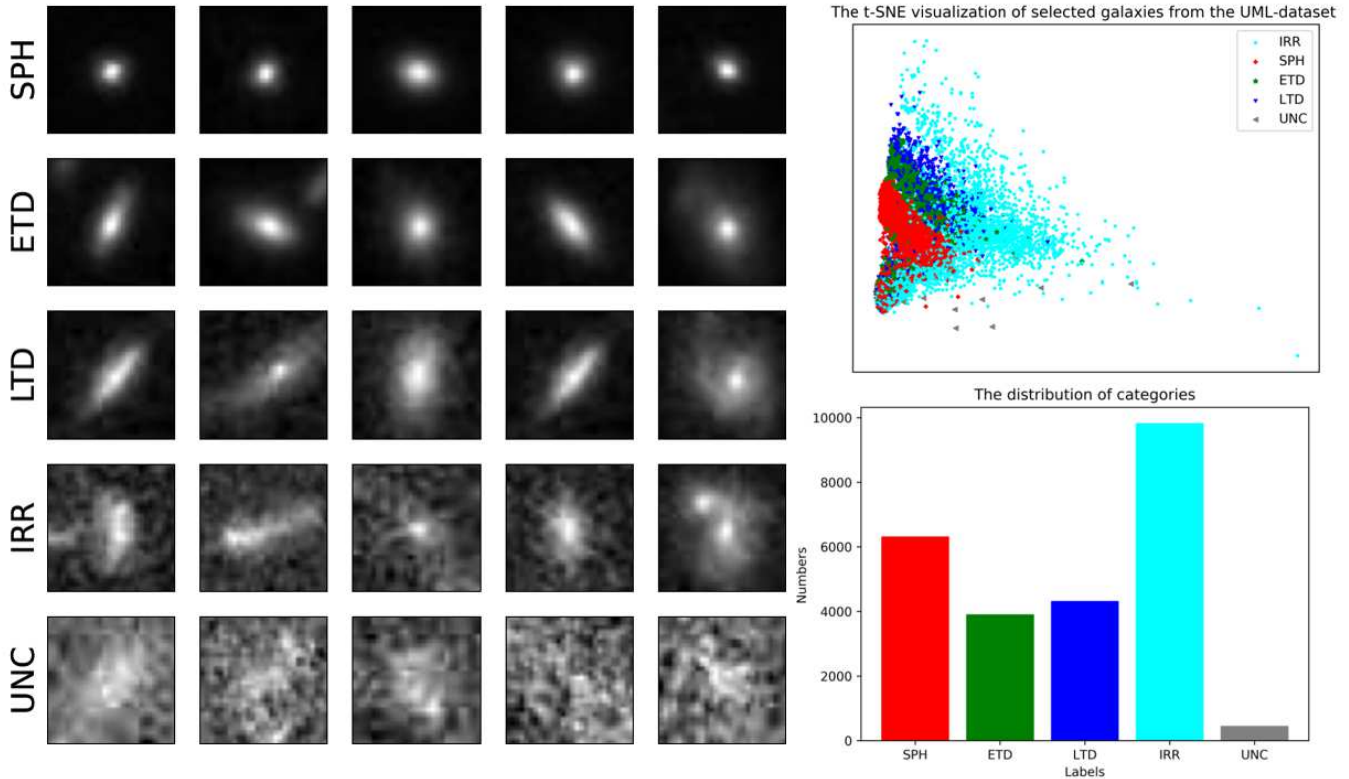}
\caption{An overview of the UML-dataset. Left: Cutouts of galaxies selected from different categories, i.e., SPH, ETD, LTD, IRR, and UNC  galaxies, are shown from top to bottom. Right: The t-SNE visualization
of 6000 randomly selected galaxies with different morphologies (top) and the corresponding counting distribution (bottom) in the UML-dataset are given. 
}
\label{uml-data}
\end{figure*}

To accomplish a well-classified dataset, a UML method, the combination of the CAE and the bagging-based multi-clustering method, is proposed in our previous work \citep{2022AJ....163...86Z}.
It has been applied to $47,149$ galaxies with $H<24.5$ extracted from five CANDLES fields. In this method, only galaxies with consistent clustering results from three clustering algorithms were defined as ``well-clustered sample'' and then ``well classified'' into different morphological types (see \citealt{2022AJ....163...86Z} for more details). As a result, $24,900$ galaxies ($\sim 53\%$) are well classified, without any pre-labeled galaxies, yielding the UML-dataset. In this dataset, galaxies are classified into five categories, including $6,335$ spheroid (SPH), $3,916$ early-type disk (ETD), $4,333$ late-type disk (LTD), $9,851$ irregular (IRR), and $465$ unclassifiable (UNC) galaxies. It provides enough samples to learn the key features corresponding to each category. The overview of UML-dataset is illustrated in Figure~\ref{uml-data}, including the example stamps, the t-Distributed Stochastic Neighbor Embedding (t-SNE) visualization, and the counting distribution of the five categories.
The t-SNE is a technique that visualizes the high dimensional data by giving each data point a location in a two or three-dimensional map \citep{van2008visualizing}.

However, the remaining $22,249$ galaxies ($\sim 47\%$; hereafter the UML remaining dataset) were eliminated in \cite{2022AJ....163...86Z} due to the disputed votes from different clustering algorithms. In order to complete the morphological classification of galaxies, the UML-dataset is considered as the training set for the downstream SML methods. That is, by training the SML models on the UML-dataset, we obtain algorithms that can give the morphological classification for the rest of the sample.

\begin{figure*}[htb!]
\centering
\includegraphics[width=\textwidth]{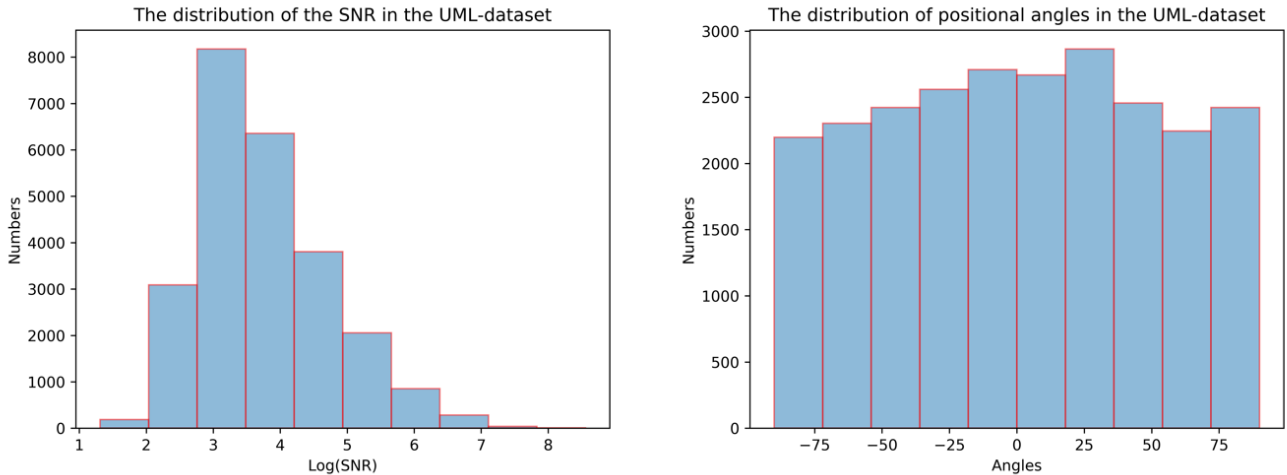}
\caption{Distributions of the image SNRs (left) and positional angles (right) of galaxies in the UML-dataset. }
\label{RN_angle}
\end{figure*}

In general, the performances of algorithms are affected by the signal-to-noise ratios (SNRs) of images and the \edit1{orientations of galaxies}. 
SNR is defined as the ratio between the flux and the corresponding uncertainty in H-band, which are extracted from \cite{2014ApJS..214...24S}. In this work, we use the astronomical definition of position angles (PA) to approximately represent the orientation of galaxies for simplicity. By assuming a single elliptical S{\'e}rsic model of light profile (central symmetric) for galaxies, PA describes the direction of the major axis of the assumed elliptical profile, which is measured with GALFIT by \cite{2014ApJ...788...28V}.
As shown in Figure~\ref{RN_angle}, due to the large sample size, galaxies in the UML-dataset have a relatively uniform PA distribution with small fluctuation. However, the distribution of image SNRs is not as uniform as that of PAs and exhibits a peak at $\sim 1000$. Problem arising from this uneven SNR distribution will be further discussed and be solved in Section \ref{subsec:noise_red}.
It is noteworthy that the astronomically defined PA ranging from -90$^\circ$ to $90^\circ $ cannot fully constrain the orientations of galaxies due to the fact that the light profiles and backgrounds of galaxies in observations are not perfectly central symmetric.
Thus, unlike the model profile of galaxies, the raw image from observation and its rotated image are not exactly the same. The standard CNN models are not rotation invariant which might recognize them as different types.
If the rotation features are not well learned by the machine, the algorithm might give different classifications for images of the same galaxy before and after rotating the image by a certain degree. However, It is expected that the classification results should not be affected by how galaxy images are rotated.
Instead of feeding in more images in the training phase by artificially rotating to solve this problem, we propose an APCT in the pre-process phase as described in Section \ref{subsec:apct}.

\section{The APCT and Other Pre-processing Strategies}\label{sec:data_pre}
The pre-processing strategies are also important for the SML algorithms.
In this section, the proposed adaptive polar coordinate
transformation and other pre-processing strategies are introduced.

\subsection{The APCT}
\label{subsec:apct}

The morphological classification of galaxies should be rotationally invariant, which means that the result of morphological classification should not change regardless of how the raw image of one galaxy is rotated. CNN extracts the local translation-invariant features effectively, which is widely used to learn the key morphological features of the galaxies.
However, the standard CNN models have poor robustness to rotations of images (see \citealt{cheng2016learning,cheng2018learning,cabrera2017deep,chen2018rotation,reyes2018enhanced,yao2019rotation}). Existing SML algorithms, such as the CNN-based neural networks, are affected by the rotational angle so that they might not recognize galaxies with the same morphological type after rotation. The performance of the SML methods will be affected, especially when the training set has an uneven distribution of orientations.

Data augmentation and conventional polar coordinate transformation are both strategies to overcome the problem of rotation invariance.
Data augmentation is a common treatment that generates extra images by rotating the raw images to make the angles evenly distributed in the training set.
Another alternative is the polar coordinate transformation that transforms the rotation of the raw images into the translation of the new images, as well as the potential features.
By combining the polar coordinate transformation with the CNN, which is translation-invariant, one obtains a rotation-invariant algorithm.
For example, the polar coordinate transformation is used to transform circle-like features in the CT Images of vulnerable plaques into line-like features \citep{Liu2019}.

\begin{figure*}[htb!]
\centering
\includegraphics[width=0.65\textwidth]{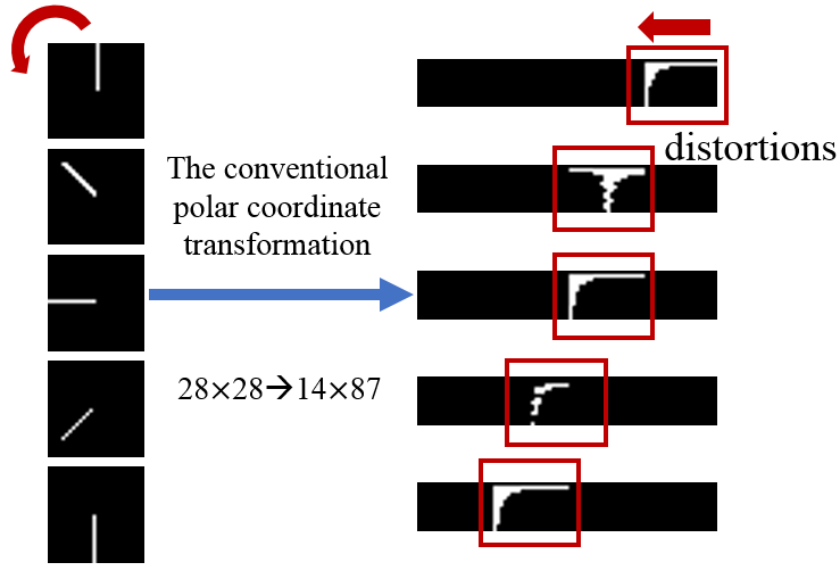}
\caption{Examples of the conventional polar coordinate transformation. It shows that the conventional polar coordinate transformation can not transform the rotations of the raw images into perfect translations of the new images.}
\label{distort}
\end{figure*}

\begin{figure*}[htb!]
\centering
\includegraphics[width=0.8\textwidth]{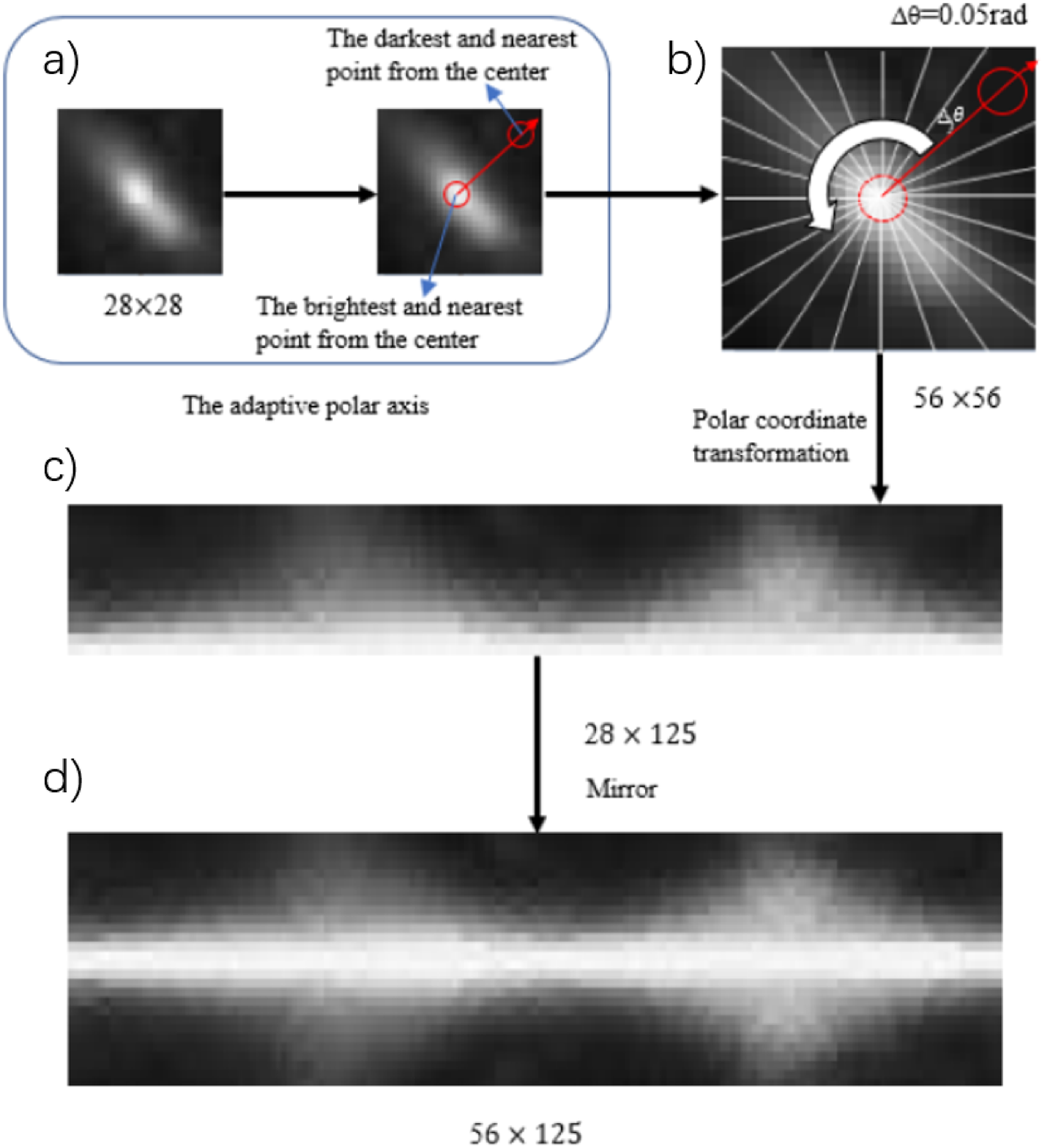}
\caption{An flow chart of the pre-process phase before feeding the network, including adaptive polar coordinate transformation, and mirroring.}
\label{polar}
\end{figure*}

To yield an efficient data pre-processing method for galaxy images, we adopt the polar coordinate transformation rather than data augmentation in the pre-processing phase of our rotationally invariant SML method. However, when the pixel values in the orthogonal coordinate are assigned to the new pixels in the polar coordinate, distortion may happen since the images are constructed by discrete pixels. As also shown in Figure~\ref{distort}, if using the conventional polar coordinate transformation, rotations of the raw images may not be transformed into translations of the new images perfectly.  Thus, in this section, we propose the APCT to solve this problem.
To help clarification, an overview of the pre-process phase is shown in Figure~\ref{polar}, while the improvements are summarized as the following three aspects.

(1)  The conventional polar coordinate transformation uses a fixed polar axis.
In that case, as the raw images rotate, there are horizontal translations in the transformed images. Different from the conventional polar coordinate transformation, the proposed APCT uses a polar axis that is invariant under rotation instead of a fixed one. In this approach, The positions of pixels with the maximum and minimum flux values in the images are selected as the brightest and darkest points, respectively. When there is more than one pixel with the minimum/maximum fluxes, we choose the ones with the smallest distance to the image centers.
The axis from the brightest point to the darkest point is taken as the polar axis of the polar coordinate system making the polar axis of the processed images rotationally invariant, as shown in panel a) of Figure \ref{polar}.

(2) For images with smaller sizes, such as $28\times28$, the polar coordinate transformation will lead to distortions, as shown in Figure \ref{distort}. We find that the distortion will be reduced if the images are enlarged.
Therefore, to reduce the distortion caused by the polar coordinate transformation,
the raw images are resized from $28\times28$ to a larger size, say $56\times56$, before the APCT. Once the polar axis is chosen, as shown in panel a) of Figure \ref{polar}, the axis is rotated counterclockwise at a unit of $0.05$ rad each time.
For each discreet rotation, the axis passes through many pixels of the raw images.
By stacking the pixels along this rotating axis while rotating, one obtains a new image with a size of $28\times 125$ with $125$ equaling $2\pi/0.05$ rounded, as shown in panels b) and c) of Figure \ref{polar}.
$28$ here is the radius equaling half of the width or height of the images after the aforementioned resizing. Pixels along the rotating axis within the coverage of images are remained, while those outside the coverage or with missing fluxes are set to be $0$. In the transformation, raw pixels will be re-sampled yielding new images with larger sizes.

(3) To highlight morphological features, images are mirrored,
resulting in images with a size of $56 \times 125$ before feeding to the algorithms, as shown in panel d) of Figure \ref{polar}.

\begin{figure*}[htb!]
\centering
\includegraphics[width=\textwidth]{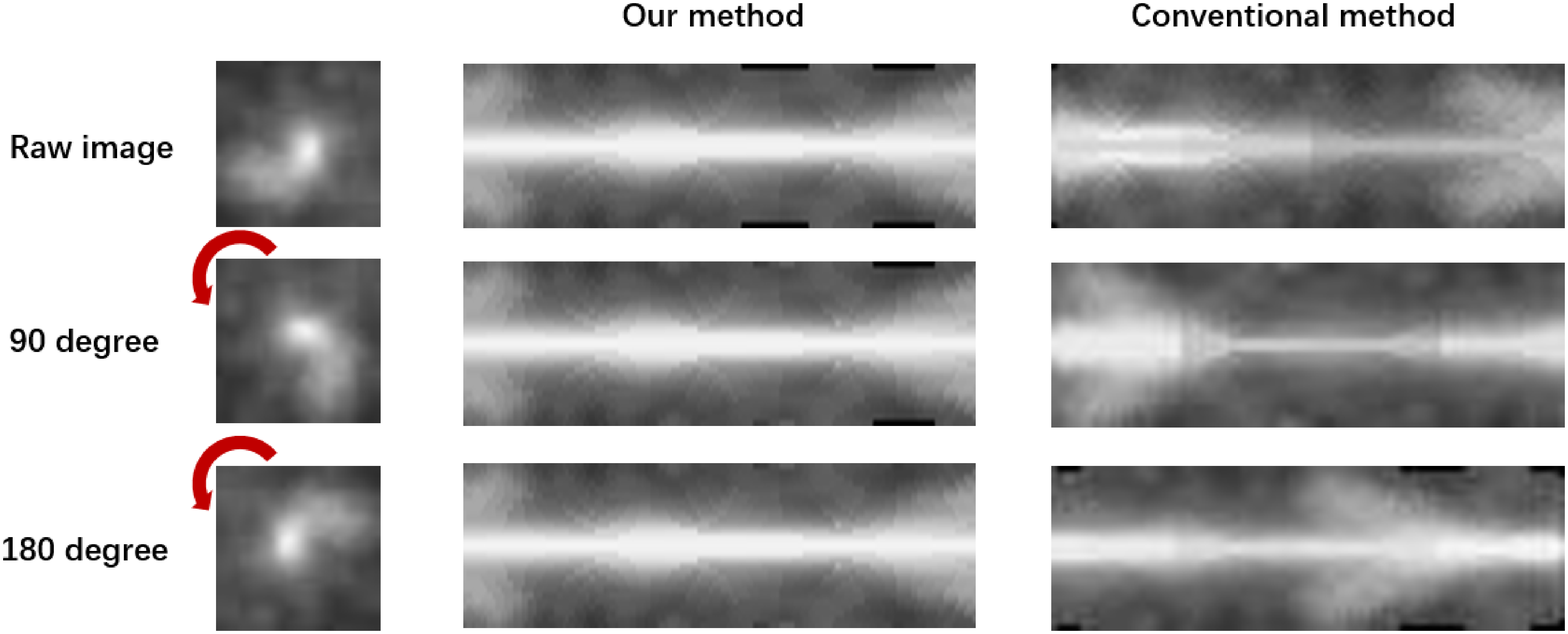}
\caption{A comparison between the conventional and the adaptive polar coordinate transformation. Our method (i.e., APCT) keeps the resulting images almost invariant regardless of how the raw images are rotated. }
\label{comp}
\end{figure*}

A comparison between the conventional polar coordinate transformation and the APCT is given in Figure~\ref{comp}. It shows that the APCT is almost rotationally invariant. For example, a simple statistic shows that, for the majority of the images after the APCT, the mean pixel-value differences before and after the rotation are smaller than $1$, which is small enough and can be ignored.

\subsection{Noise Reduction by Convolutional Autoencoder}
\label{subsec:noise_red}

The CNNs are sensitive to noises (see \citealt{liu2020networks,nazare2017deep}),
therefore, the distribution of image SNRs in the training set also affects the performance of the SML algorithms.
For example, if the training set has an uneven distribution of the SNRs, say most of the samples in the training set have high SNRs, the algorithm might fail on the test set consisting of samples with low SNRs. It is because in images with low SNRs,
the noises will break features that are learned by the neural networks leading to misclassifications.
As shown in the left panel of Figure \ref{RN_angle}, the distribution of image SNRs of the UML-dataset is not uniform, thus additional procedure in the image pre-process or the training algorithms, such as enhancing the image quality of low-SNR images or assigning more weights to low-SNR images in the training, is required. As an effective implementation method of the former procedure (i.e., enhancing the poor quality of low-SNR images), noise reduction is known to be a useful method to overcome the problem arising from the uneven distribution of SNR (see \citealt{nazare2017deep}). In this work, we adopt the CAE method to reduce the noises \citep{7434593}.

\begin{figure*}[htb!]
\centering
\includegraphics[width=\textwidth]{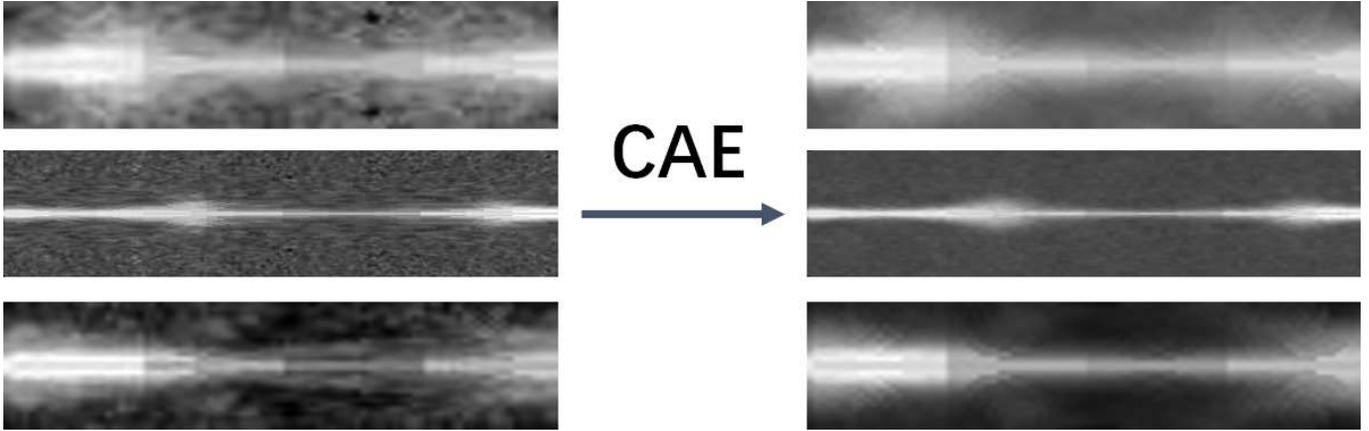}
\caption{Comparisons between the input (left) and reconstructed (right) images in the noise reduction by using the CAE.
The reconstructed images not only maintain the main morphological features but also remove the redundant pixels, and thus reduce the noises.}
\label{CAE}
\end{figure*}

In this approach, the images are encoded by the convolutional and subsampling layers and decoded by the deconvolutional and upsampling layers generating the reconstructed images. The parameters of the CAE are trained to minimize the mean square error of pixels between the input and reconstructed images. We show some examples of comparisons between the input and reconstructed images in Figure~\ref{CAE}. The reconstructed images not only maintain the main morphological features but also remove the redundant pixels, and thus reduce the noises.

The pre-processing strategies of APCT and noise reduction make the SML algorithms not affected by the distributions of PAs and SNRs of galaxies in the training set. In other words, by applying the proposed pre-process strategies, we reduce the dependence of the training set on the galaxy PAs and image SNRs.

\section{The SML Models and the Experiment Settings }\label{sec:method}

In order to complete the entire task of automatic morphological classification of galaxies and, at the same time, give the morphological types for the UML remaining dataset, we train three widely used SML algorithms on the UML-dataset. In this section, we give a brief review of the three SML methods and the experiment settings.

\subsection{A Brief Review of three SML Models}

\begin{figure*}[htb!]
\centering
\includegraphics[width=\textwidth]{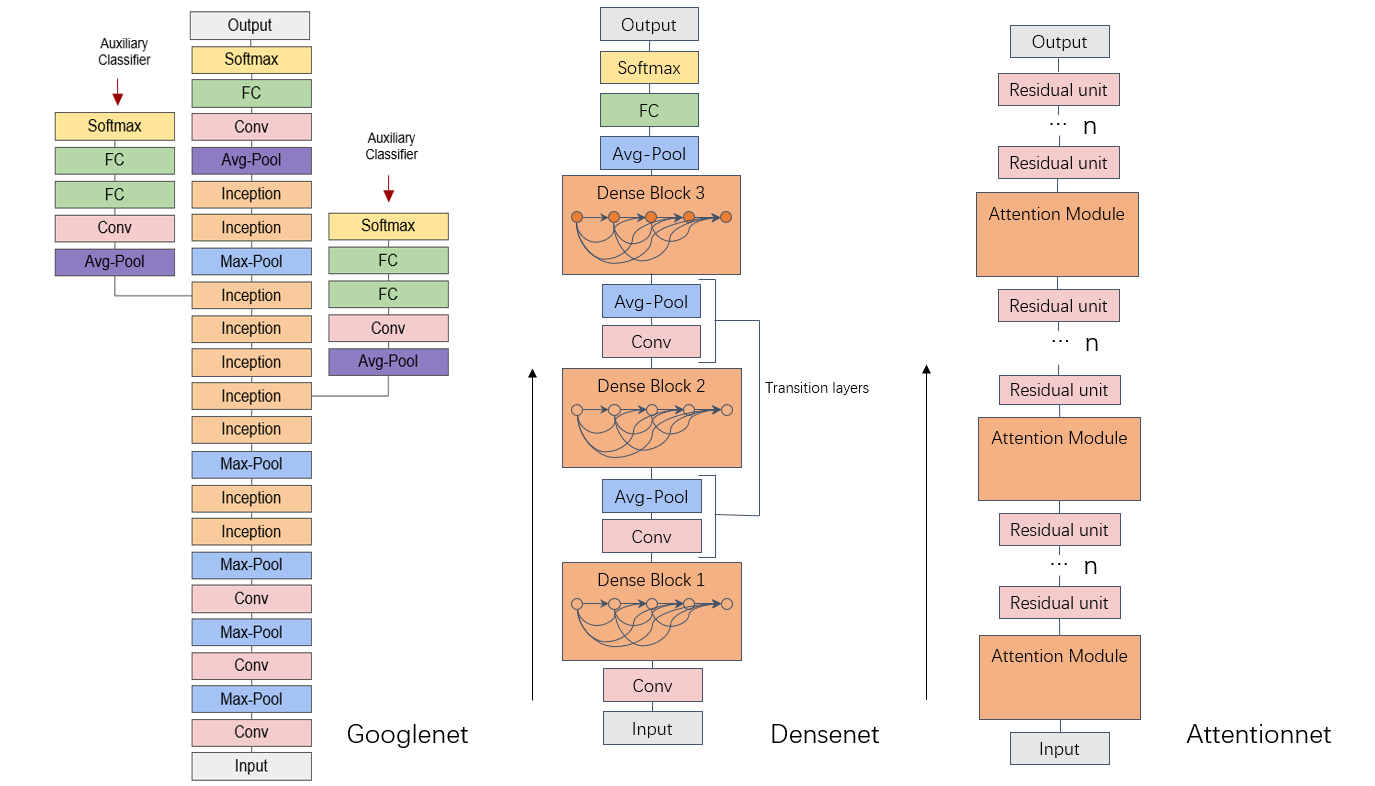}
\caption{An overview of the three SML networks.}
\label{networks}
\end{figure*}

In this work, we test three SML models, namely the GoogLeNet  (see \citealt{7298594}), the DenseNet121 (see \citealt{huang2017densely}), and the attention56 network  (see \citealt{wang2017residual}). The one with the best performance on the validation set will be selected as our fiducial model.
Figure~\ref{networks} gives an overview of the three neural networks. The brief introductions are described below.

The GoogLeNet is a deep learning neural network proposed by
Christian Szegedy in 2014 (see \citealt{7298594}).
The GoogLeNet uses $9$ inception modules stacked one over the other. In the inception module, convolutions with kernel sizes of $1\times1$, $3\times3$, and $5\times5$ and max pooling with a kernel size of $3\times3$
are applied in a parallel way, making it cover a bigger area of the images and, at the same time, keep a fine resolution for small areas.

The DenseNet121 is one of the derivative versions of the densely connected convolutional network (DenseNet) (see \citealt{huang2017densely}). In a DenseNet, each layer is connected to every other layer.
That is, the outputs of all the preceding layers are used as inputs of the subsequent layers. Unlike the resnet, which uses an additive method that merges the previous layer with the future layer, the DenseNet concatenates the outputs of the previous layers with the future layers. The number ``121'' denotes the number of layers in the neural network.

The attention56 network is one of the derivative versions of the residual attention network (see \citealt{wang2017residual}). The residual attention network is built by stacking attention modules.
The attention modules can generate attention-aware features and the attention-aware features from different modules change adaptively as layers go deeper.
Inside each attention module, a bottom-up top-down feedforward structure is used to unfold the feedforward and feedback attention process into a single feedforward process (see \citealt{wang2017residual}).
The number ``56'' denotes the number of layers in the neural network.

\subsection{Experiment Setting}

One of the main aims of this work is to train the SML models on the UML-dataset with $24,900$ galaxies
and then label the UML remaining dataset with $22,249$ galaxies that cannot be well classified by the UML method proposed by \cite{2022AJ....163...86Z} in the five CANDELS fields.
In principle, all the UML-dataset should be trained in order to make the algorithms learn all the information contained in the training set.
However, to avoid overfitting, the UML-dataset should be split into subsamples of training and validation sets usually with a ratio of $7:3$ or $8:2$. Here, instead of adopting a fixed ratio, the UML-dataset is randomly split into training and validation sets containing $22,000$ and $2,900$ galaxies, respectively, as shown in Table~\ref{tab_2.1}.
To ensure that there is no overfitting, we repeat the random splitting of the UML-dataset several times to generate different combinations of the training and validation sets and perform the training and validation on these sets respectively. The result indeed shows that there is no overfitting in our training.

\begin{deluxetable*}{ccccccc}[htb!]
\setlength{\tabcolsep}{5mm}
\tablecaption{The numbers of galaxies in the training and validation sets. \label{tab_2.1}}
\tablehead{\colhead{Number of images} &
\colhead{SPH} &
\colhead{ETD} &
\colhead{LTD} &
\colhead{IRR} &
\colhead{UNC} &
\colhead{ALL}
}
\startdata
Training set & $5588$ & $3436$ & $3837$ & $8727$ & $412$ & $22000$ \\
Validation set & $747$ & $480$ & $496$ & $1124$ & $53$ & $2900$ \\
\enddata
\end{deluxetable*}

In the training phase of the SML algorithms, the batch size is $32$, the learning rate for $0.0001$, and the maximum training epoch is $500$.

\section{Results and Analysis}\label{sec:result}
In this section, we show the result of the proposed rotationally invariant SML method. The performance is mainly evaluated by the accuracy of the validation set and the t-SNE visualization on the UML remaining dataset. We will also demonstrate the effectiveness of the pre-process strategy of the APCT.

\subsection{Performances of three models}
In this section, we show the performances of three SML models.
The morphological labels from the UML-dataset are regarded as the real labels,
whereas the predicted labels are given by the SML methods.
The overall accuracy is calculated by $N_{\rm correct}/N_{\rm total}$ where $N_{\rm correct}$ is the number of correctly labeled galaxies and $N_{\rm total}$ is the total number of galaxies.
The overall accuracies of the three models on the validation sets are listed in Table \ref{tab:acc}.
It shows that all three models have high accuracy ($>90\%$), while the GoogLeNet has the highest accuracy of $95.1\%$. Table \ref{tab:acc} also shows that the overall accuracy of GoogLeNet on the validation set is improved by $\sim 5.2\%$ after the noise reduction, proving the effectiveness of noise reduction by the CAE.

\begin{deluxetable*}{cccccc}[htb!]
\setlength{\tabcolsep}{5mm}
\tablecaption{Overall accuracy of the deep learning models. \label{tab:acc}}
\tablehead{\colhead{Model} &
\colhead{Accuracy of the training set} &
\colhead{Accuracy of the validation set} &
}
\startdata
GoogLeNet &  $ 100.0\% $ & $ 95.1\% $ \\
GoogLeNet (Without Noise Reduction) &  $ 100.0\% $ & $ 89.9\% $ \\
DenseNet121 & $ 100.0\% $ & $ 94.3\% $ \\
DenseNet121 (Without Noise Reduction) &  $ 100.0\% $ & $ 89.9\% $ \\
Attention56 & $ 100.0\% $ & $ 93.2\%$ \\
Attention56 (Without Noise Reduction) &  $ 100.0\% $ & $ 89.6\% $ \\
\enddata
\end{deluxetable*}

\begin{figure*}[htb!]
\centering
\includegraphics[width=\textwidth]{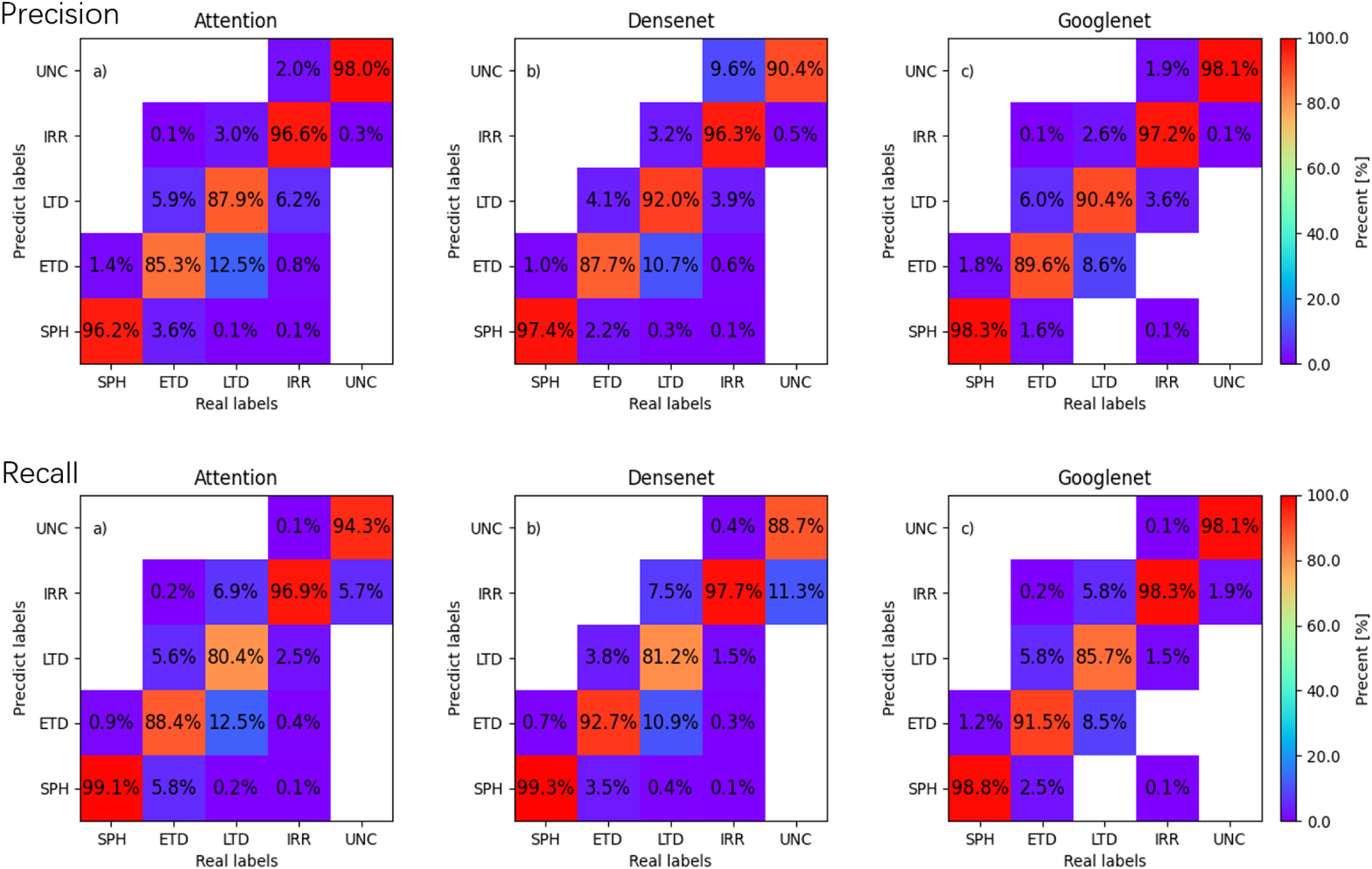}
\caption{The precision (top) and the recall (bottom) of each morphological types in the validation set for the three SML models, i.e., Attention56 (left), DenseNet121 (middle), and GoogLeNet (right).}
\label{fig_sml}
\end{figure*}

Figure~\ref{fig_sml} shows the precision and recall of the predictions. The precision is defined by $N_{\rm predict-correct}/N_{\rm predict-total}$, where $N_{\rm predict-total}$ is the total number of this morphological type predicted by the SML model and the $N_{\rm predict-correct}$ is the number of correct predictions in line with the labels from the UML-dataset. The recall is defined by $N_{\rm predict-correct}/N_{\rm real-total}$, where $N_{\rm real-total}$ is the total number of this morphological type from the UML-dataset.
It shows that although the DenseNet121 model has higher precision in the LTD category, GoogLeNet performs much better in the other four categories making it the best model. Therefore, in the following analysis, GoogLeNet is chosen to be the representative and fiducial SML model.

\subsection{The Effectiveness of the APCT }

In this section, we demonstrate the effectiveness of the APCT. Without loss of generality, we rotate the images in the validation set at 90\arcdeg, 180\arcdeg, and 270\arcdeg, respectively, and calculate the overall accuracy for the GoogLeNet model.

\begin{figure*}[htpb]
\centering
\includegraphics[width=1.0\textwidth]{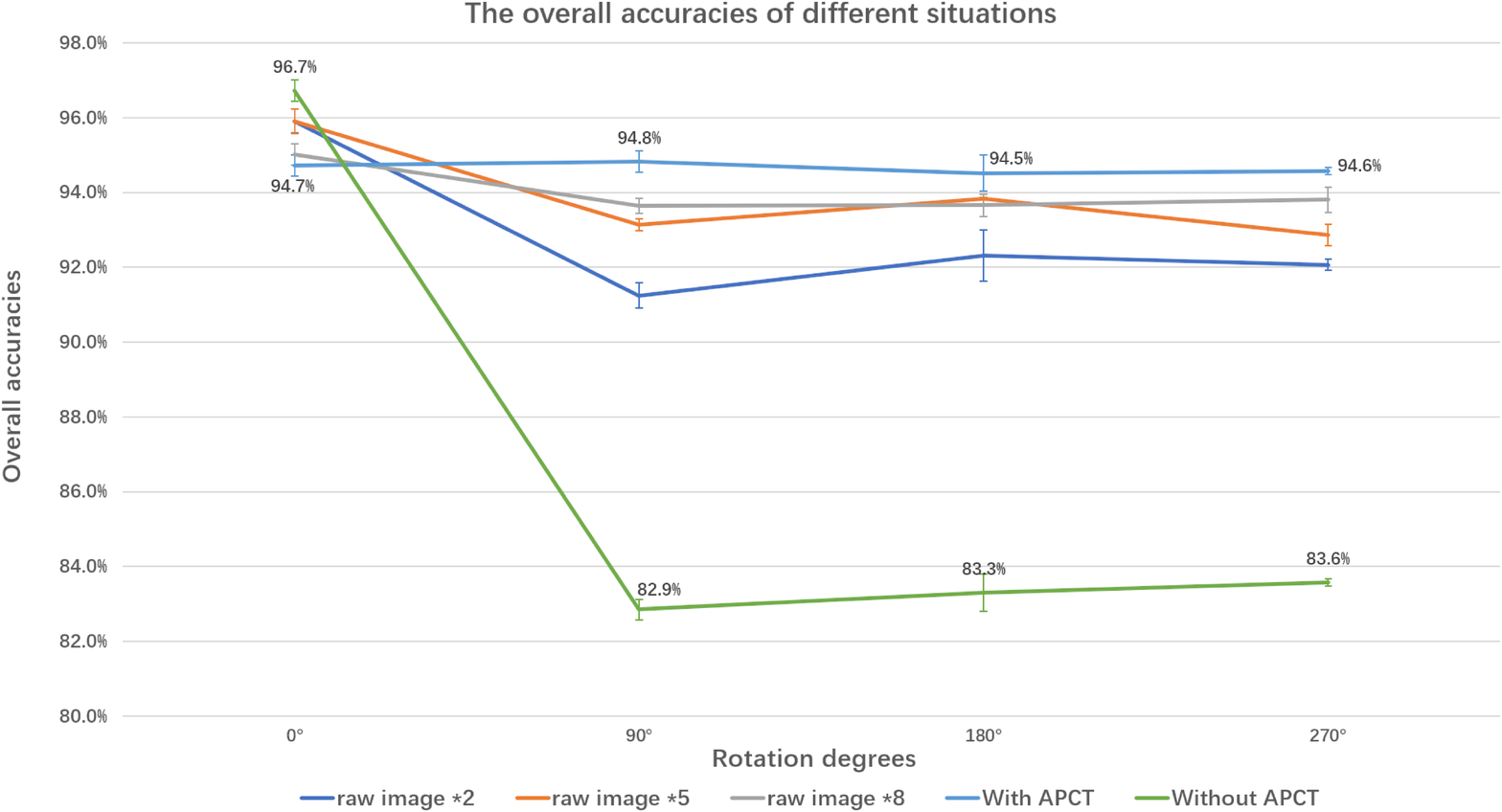}
\caption{The accuracy of the GoogLeNet with and without APCT on the validation sets as a function of rotational angles.}
\label{fig_rotate}
\end{figure*}

The accuracy as a function of rotational angles for the GoogLeNet model applied to the images with and without APCT is presented in Figure~\ref{fig_rotate}, which shows that under the APCT, the overall accuracy is nearly unchanged after rotations compared to that of the original, rotation-free validation set. However, the overall accuracy of the models using the raw images after rotations but without APCT decreases significantly (by about 13\%).
Given that the un-rotated validation set is randomly selected from the UML-dataset, there is little difference in the distributions of angles between the training and validation sets. Therefore, the un-rotated validation set has a high overall accuracy without the APCT.
However, after rotation, the difference in the angle distributions between the training and validation sets occurs. As a result, the overall accuracy decrease obviously from over $95\%$ to about $83\%$, suggesting that the existing SML methods have poor robustness to rotations of images.

Moreover, Table~\ref{three_rotate} shows the overall accuracy of the original and rotated validation sets of the three SML models. It shows that, after rotation, the accuracy is nearly unchanged (with the largest difference of only 1.5\%) compared to that of the un-rotated validation set regardless of the SML models considered.

In practical application, the model needs to label galaxies that have never been met before. The distribution of angles of the new dataset might have a large difference from that of the training set.
Therefore, the SML method with good robustness to rotations of images is preferred. Figure~\ref{fig_rotate} and Table \ref{three_rotate} demonstrate that the APCT is an effective and efficient method to improve the robustness of the SML methods.

\begin{deluxetable*}{cccccc}[htbp]
\setlength{\tabcolsep}{5mm}
\tablecaption{The accuracy of three SML models with and without APCT on the validation sets rotated by
different angles. \label{three_rotate}}
\tablehead{\colhead{Model} &
\colhead{Without rotation} &
\colhead{90\arcdeg} &
\colhead{180\arcdeg} &
\colhead{270\arcdeg} &
}
\startdata
GoogLeNet &  $ 95.1\% $ & $ 94.7\% $ & $ 94.5\% $ & $ 94.6\% $\\
GoogLeNet(Without APCT) &  $ 96.7\% $ & $ 82.9\% $ & $ 83.3\% $ & $ 83.6\% $\\
DenseNet121 & $ 94.3\% $ & $ 93.6\% $ & $ 93.4\% $ & $ 93.9\% $\\
DenseNet121(Without APCT) & $ 97.2\% $ & $ 82.3\% $ & $ 82.6\% $ & $ 83.4\% $\\
Attention56 & $ 93.2\% $ & $ 92.0\% $ & $ 91.7\% $ & $ 92.9\% $\\
Attention56(Without APCT) & $ 96.0\% $ & $ 82.4\% $ & $ 76.3\% $ & $ 79.6\% $\\
\enddata
\end{deluxetable*}

\subsection{The t-SNE visualization of the UML remaining dataset}

Given that the intrinsic morphological types of the galaxies in the five CANDELS fields are missing, we can not directly evaluate the performance of the SML methods by calculating indicators such as overall accuracy.
In this section, we use the t-SNE visualization graph to evaluate the morphological classification result of the UML remaining dataset. The t-SNE is a useful method to map the subsamples from a high-dimensional feature space to a 2-dimensional compressed feature space.
By using this method, one can check the classification result intuitively, especially when the real labels are missing.

\begin{figure*}[htb!]
\centering
\includegraphics[width=\textwidth]{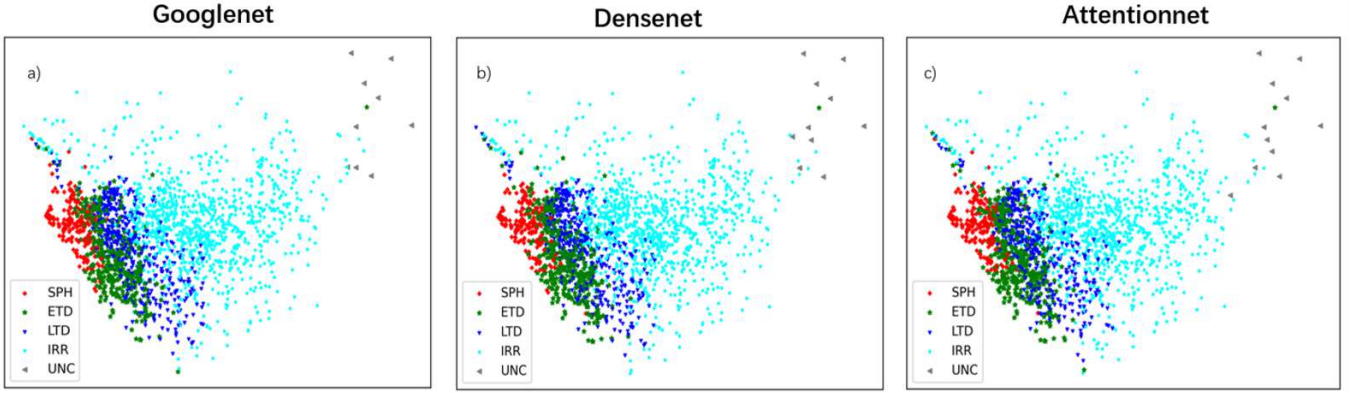}
\caption{The t-SNE visualization graphs of $2,000$ randomly selected images that are labeled by the three SML methods. There is no obvious evidence that one model has an advantage over the two.}
\label{fig02}
\end{figure*}

The morphological types of the UML remaining dataset with $22,249$ galaxies are predicted by our SML method.
Here, to verify the effectiveness of the morphological classifications, we randomly select $2,000$ labeled galaxies from the dataset and give the t-SNE visualization graphs based on the raw images.
Figure \ref{fig02} shows the t-SNE visualization graphs of the $2,000$ galaxies labeled by three methods.
It shows that the SML models give the labeled galaxies with clear boundaries that prove the effectiveness of the proposed method.
From the t-SNE visualization graphs, there is no obvious evidence that one model has an advantage over the other two models.

\begin{figure*}[htb!]
\centering
\includegraphics[width=\textwidth]{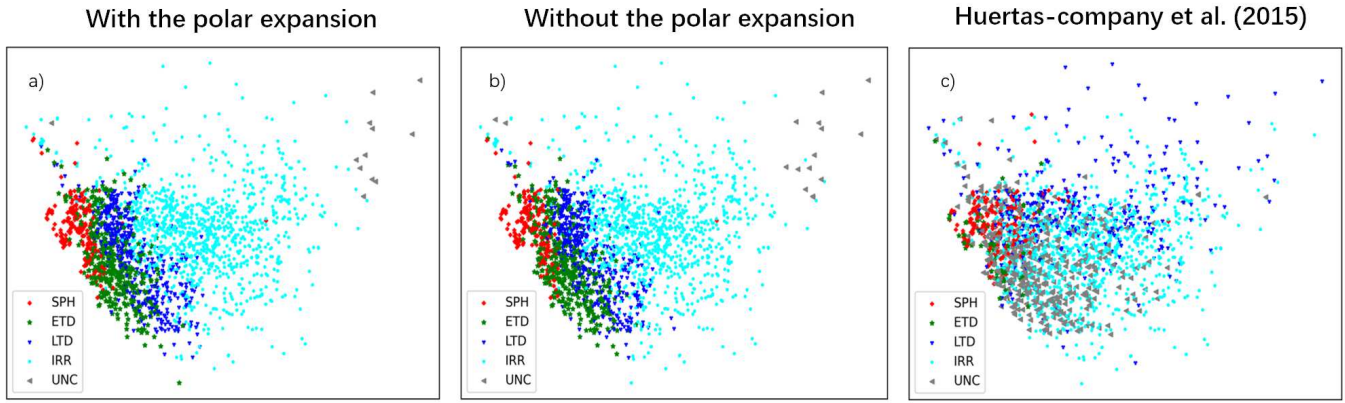}
\caption{The t-SNE visualization graphs of $2,000$ randomly selected images that are labeled by different results. Image a) is the results of the GoogLeNet model with APCT; Image b) is the results of the GoogLeNet model without APCT; Image c) is the results of \cite{2015ApJS..221....8H}. It shows that both SML models give the labeled galaxies with clear boundaries in the t-SNE visualization graphs, proving the effectiveness of the proposed method.}
\label{fig03}
\end{figure*}

To further evaluate the effectiveness of the proposed rotationally invariant SML method, we give Figure \ref{fig03} that shows the t-SNE visualization graphs between our SML methods with/without the APCT and the SML results of \cite{2015ApJS..221....8H}. The morphological type of \cite{2015ApJS..221....8H} is determined by five parameters, $f_{\rm spheroid}$, $f_{\rm disk}$, $f_{\rm irr}$, $f_{\rm PS}$ and $f_{\rm Unc}$. The definition is shown as follows:
\begin{enumerate}
\item Spheroids (SPH): $f_{\rm spheroid} > 2/3$, $f_{\rm disk} < 2/3$, and $f_{\rm irr} < 0.1$ ;
\item Early-type Disks (ETD): $f_{\rm spheroid} > 2/3$, $f_{\rm disk} > 2/3$, and $f_{\rm irr} < 0.1$ ;
\item Late-type  Disks (LTD): $f_{\rm spheroid} < 2/3$, $f_{\rm disk} > 2/3$, and $f_{\rm irr} < 0.1$;
\item Irregulars (IRR): $f_{\rm spheroid} < 2/3$ and $f_{\rm irr} > 0.1$.
\item Unclassifiable (UNC): the remaining sources.
\end{enumerate}
\normalsize
Galaxies can be matched by the object ID from the 3D-HST catalogs of \cite{2014ApJS..214...24S}.

It shows that the SML models trained on the UML-dataset give the labeled galaxies with clear boundaries in the t-SNE visualization graphs.
Among $22,249$ galaxies in the UML remaining dataset, $17,793$ of them ($\sim 80\%$) are labeled consistently by the GoogLeNet with and without APCT.
Like the result shown in Figure \ref{fig02} where the differences between the three models are not obvious, in this t-SNE visualization graph (Figure \ref{fig03}) the difference between the classification result with and without APCT is also not obvious. One plausible reason for this situation is that the t-SNE graph is not a quantitative tool and thus not good at distinguishing small differences in accuracy, like difference $\lesssim 10\%$ (e.g., Figure \ref{fig_rotate}).

\begin{figure*}[htb!]
\centering
\includegraphics[scale=0.5]{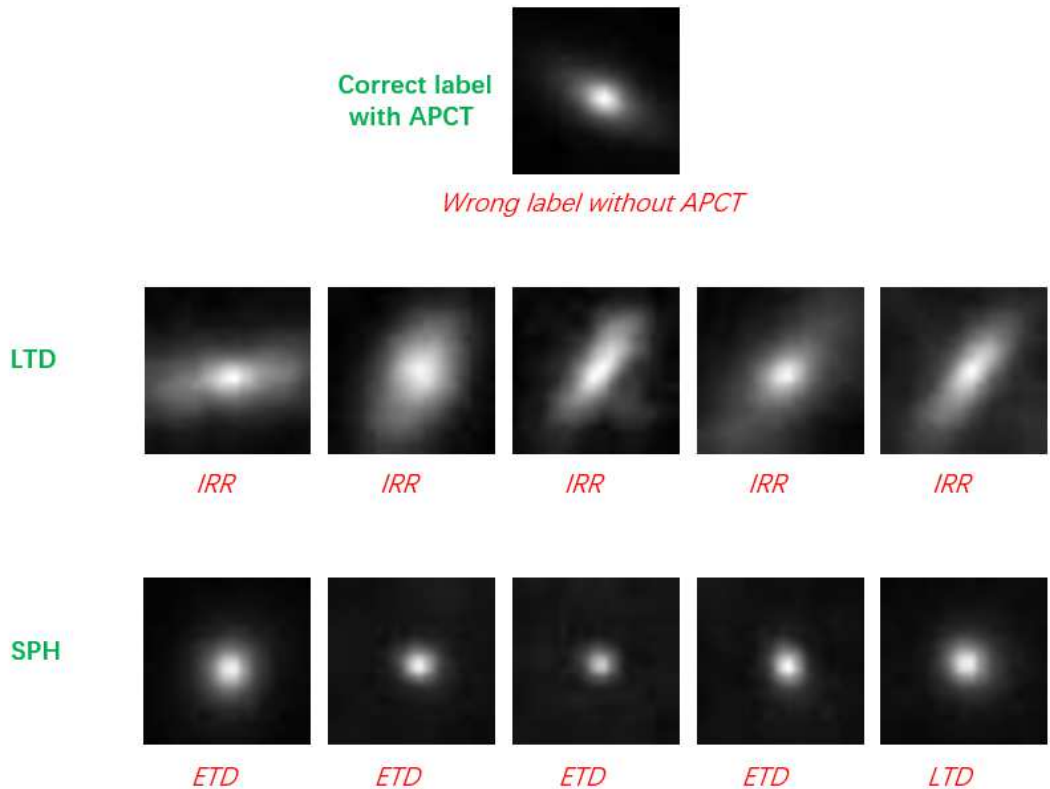}
\caption{Examples of galaxies that are correctly labeled by the rotationally invariant SML model with APCT but incorrectly labeled by the same model without APCT. These galaxies are firstly randomly selected from galaxies with disputed labels by the SML models with and without APCT and then verified by visual inspection.}
\label{examples}
\end{figure*}

Moreover, in Figure \ref{examples} we show some galaxies selected from the UML remaining dataset that are correctly labeled by the rotationally invariant SML model with APCT but incorrectly labeled by the same model without APCT.
These galaxies are firstly randomly selected from the collections of $4,456$ galaxies that are inconsistently labeled by the GoogLeNet with and without APCT and then verified by visual inspection. It shows that the proposed rotationally invariant SML model with APCT has better robustness to morphological types such as LTD and SPH.

\subsection{Comparisons with galaxy properties}

\begin{figure*}[htb!]
\centering
\includegraphics[width=\textwidth]{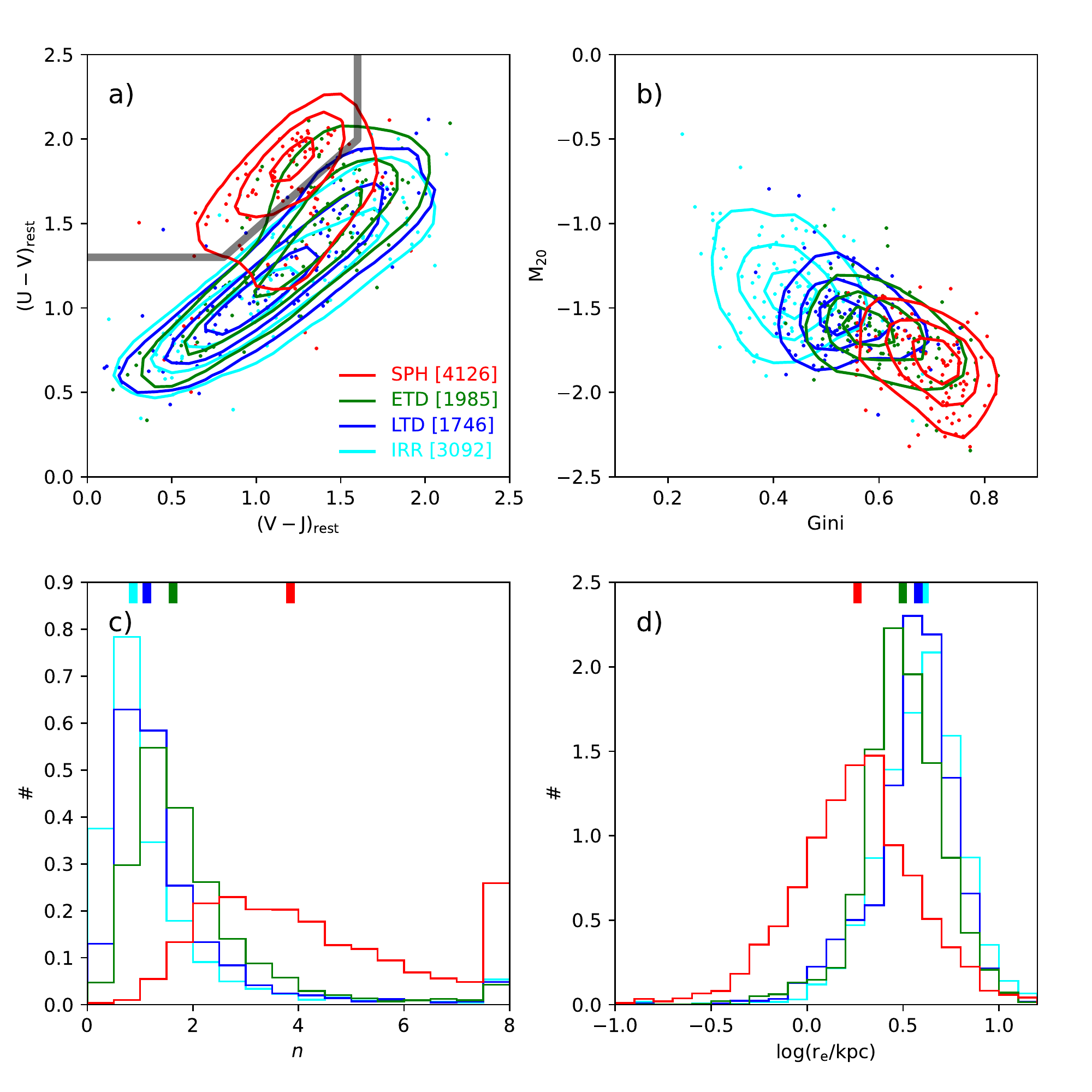}
\caption{Relationships between morphological types and other galaxy properties for massive galaxies in the UML remaining dataset. The upper panels summarize the distribution of galaxies in rest-frame $U-V$ v.s. $V-J$ color space and in the $G-M_{20}$ space. The contour levels indicate the 20\%, 50\%, and 80\% of the corresponding subclass from inner to outskirts. The data points are 100 galaxies randomly selected from each  subclass, where the total numbers are given in the corner. The bottom panels are the distribution of $n$ and $r_e$, with the median values represented as the upper bricks. The SPH, ETD, LTD, and IRR categories are represented by red, blue, green, and cyan colors, respectively.}
\label{fig_galaxy}
\end{figure*}

Similar to our previous work of \cite{2022AJ....163...86Z}, we investigate the connections of massive galaxies ($M_*> 10^{10} M_\odot$) between our morphological types and other galaxy properties in Figure \ref{fig_galaxy}. The SPH, ETD, LTD, and IRR categories are represented by red, green, blue, and cyan colors, respectively.
The upper left panel summarizes the distribution of galaxies in rest-frame $U-V$ v.s. $V-J$ color space. The wedged region defined by \cite{2009ApJ...691.1879W} represents the region dominated by quiescent galaxies and the rest of the region represents the region where star-forming galaxies reside.
The combination of the Gini coefficient and $M_{20}$ shown in the upper right panel is useful to demonstrate the disturbance and bulge strength of galaxies \citep{2004AJ....128..163L, 2020ApJ...899...85S}. The Gini coefficient can quantify the uniformity of light distribution. The higher Gini means that fluxes are more concentrated in the minority of pixels. The $M_{20}$ is the second moment of the distribution of the brightest 20\% of galaxy light, tracing the substructures in a galaxy.
The bottom panels are the distribution of the S\'{e}rsic indices $n$ and effective radii $r_e$, respectively. The detailed definitions of these parameters have been introduced in our previous work of \cite{2022AJ....163...86Z}.

In general, the galaxy properties agree well with the expected sequence of our morphological types from IRR to SPH. Following the sequence in Figure \ref{fig_galaxy}, Panel a) shows that galaxies move from the star-forming region to the quiescent region; Panel b) illustrates galaxies changing from more disturbed to more concentrated (high Gini coefficient for SPH category); Panel c) shows that galaxies tend to be more bulge dominated; Panel d) implies that galaxies become smaller in size. In \cite{2022AJ....163...86Z} we have demonstrated that the majority of the UNC sources have very low SNRs (e.g., images in the bottom row of the left panel of Figure \ref{uml-data}). For this reason, we do not carry out a similar analysis for these UNC sources.

\section{Conclusions and outlooks}
\label{sec:sum_outlooks}

The main aim of this series of works is to provide an automatic morphological classification method for galaxies.
To accomplish this goal, we implement the following studies.

(1) In \cite{2022AJ....163...86Z}, a UML method was proposed to build a well-classified UML-dataset. In that approach, the CAE is used to compress the dimension of raw image data and, at the same time, extract key morphological features.
The bagging-based multi-clustering method, a voting method, clusters galaxies with analogous characteristics into one group.
By applying this UML method to galaxies in the five CANDELS fields, we obtain a well-classified UML-dataset consisting of $24,900$ galaxies that are consistently voted by different clustering algorithms at the cost of eliminating $22,249$ disputed galaxies. The UML-dataset is generated without using any pre-labeled galaxies.

(2) In this work, we develop a rotationally invariant SML method as supplementary. In this approach, an APCT is proposed to improve the robustness of the SML method to rotations of images.
Then, the SML algorithms are trained on the UML-dataset giving a method that identifies the morphological types of new galaxies with high confidence. By applying the rotationally invariant SML method on galaxies in the five CANDELS fields,
we give the missing labels for the remaining $22,249$ disputed galaxies. The result of the proposed method in the five CANDELS fields, including the t-SNE visualization graphs showing our morphological classification result and that of \cite{2015ApJS..221....8H}, the comparison with galaxy properties showing the agreement between the galaxy properties and the sequence of our morphological types,
proves that the combination of the UML method proposed in the previous work \citep{2022AJ....163...86Z} and the rotationally invariant SML method proposed in the present work can automatically give the morphological classification of galaxies.

The framework of automatic classification of galaxy morphology developed in this series of works will be iterated and updated continuously (e.g., Dai et al., in preparation), and could be used in future deep field surveys that produce enormous amounts of photometric data, including surveys scheduled by the forthcoming Chinese Space Station Telescope.

\begin{acknowledgments}
This work is based on observations taken by the 3D-HST
Treasury Program (GO 12177 and 12328) with the NASA/
ESA HST, which is operated by the Association of Universities for Research in Astronomy, Inc., under NASA contract NAS526555. This work has made use of the Rainbow Cosmological Surveys Database, which is operated by the Centro de Astrobiología (CAB/INTA), partnered with the University of California Observatories at Santa Cruz (UCO/Lick,UCSC).
This work is supported by the Strategic Priority Research Program of Chinese Academy of Sciences (No. XDB 41000000), the National Key R\&D Program of China (2017YFA0402600),
and the China Manned Space Project with No. CMS-CSST-2021-A07. This work is also supported by the National Natural Science Foundation of China (NSFC; Nos. 11973038, 62106033, 11673004).
C.C.Z. acknowledges the support from Yunnan Youth Basic Research Projects (202001AU070020).
Y.Z.G. acknowledges the support from China Postdoctoral Science Foundation (2020M681281) and Shanghai Post-doctoral Excellence Program (2020218).
Z.S.L. acknowledges the support from China Postdoctoral Science Foundation (2021M700137).
\end{acknowledgments}

\bibliographystyle{aasjournal}
\bibliography{ref}
\end{document}